\documentclass[twocolumn,showpacs,preprintnumbers,amsmath,amssymb,prb,superscriptaddress]{revtex4}

\usepackage{graphicx,here}
\usepackage{dcolumn}
\usepackage{bm}
\usepackage{enumerate}

\begin{document}

\preprint{APS/123-QED}

\title{Spin-Driven Bond Order in a 1/5-Magnetization Plateau Phase in a Triangular Lattice Antiferromagnet CuFeO$_2$}%

\author{Taro Nakajima}
\email{E-mail address: nakajima@nsmsmac4.ph.kagu.tus.ac.jp}
\affiliation{Department of Physics, Faculty of Science, Tokyo University of Science, Tokyo 162-8601, Japan}%
\author{Noriki Terada}
\affiliation{National Institute for Materials Science, Tsukuba, Ibaraki 305-0047, Japan}
\affiliation{ISIS Facility, STFC Rutherford Appleton Laboratory, Chilton, Didcot, Oxfordshire, OX11 0QX, United Kingdom}%
\author{Setsuo Mitsuda}
\affiliation{Department of Physics, Faculty of Science, Tokyo University of Science, Tokyo 162-8601, Japan}%
\author{Robert Bewley}
\affiliation{ISIS Facility, STFC Rutherford Appleton Laboratory, Chilton, Didcot, Oxfordshire, OX11 0QX, United Kingdom}%

\begin{abstract}
We have investigated spin-wave excitations in a magnetic-field-induced 1/5-magnetization plateau phase in a triangular lattice antiferromagnet CuFeO$_2$ (CFO), by means of inelastic neutron scattering measurements under applied magnetic fields of up to 13.4 T. %
Comparing the observed spectra with the calculations in which spin-lattice coupling effects for the nearest neighbor exchange interactions are taken into account, we have determined the Hamiltonian parameters in the field-induced 1/5-plateau phase, which directly show that CFO exhibits a bond order associated with the magnetic structure in this phase. %
\end{abstract}


\maketitle
\section{INTRODUCTION}

Geometrically frustrated magnets are fertile ground for exotic spin-lattice coupling phenomena.\cite{S.Ji_PRL_ZnCr2O4,NPhys_Matsuda2007,SpinDrivenJT_PRL,OrderByDistortion_PRL} %
Because of the topology of the lattices and the intricate magnetic interactions, %
the frustrated magnets tend not to exhibit magnetically ordered states even at low temperatures. %
To relieve the geometrical spin frustration, they often exhibit `spin-driven' crystal lattice distortions, which lower the symmetry of the lattices and lift the vast ground-state degeneracy. %

From 2000s, the spin-lattice coupling phenomena have been intensively investigated using spinel compounds, for instance $A$Cr$_2$O$_4$ or $A$V$_2$O$_4$ ($A=$ Zn, Cd, Hg and Mg), which are regarded as Heisenberg antiferromagnets with pyrochlore lattices.\cite{SpinDrivenJT_PRL,OrderByDistortion_PRL,HalfMagPlateauCdCr2O4_PRL,NPhys_Matsuda2007,S.Ji_PRL_ZnCr2O4} %
The strong spin frustration on the pyrochlore lattices are relieved by cubic-to-tetragonal (or orthorhombic) structural transitions, and consequently, antiferromagnetic orderings appear at low temperatures. %
Previous theoretical studies have pointed out that the symmetry-lowering structural transitions result in bond-order states in which exchange interactions for nearest neighbor bonds in a tetrahedron are enhanced or reduced reflecting the magnetic orderings, specifically the spin correlation $\langle\mbox{\boldmath $S$}_i\cdot\mbox{\boldmath $S$}_j\rangle$ on each bond.\cite{SpinDrivenJT_PRL,OrderByDistortion_PRL} %
This phenomenon may be called `spin-driven' bond order. %
 
It is also known that the spin-lattice coupling effect also plays a crucial role in magnetic-field-induced phase transitions in Cr-spinel oxides, which exhibit 1/2-magnetization plateau states under applied magnetic field.\cite{HalfMagPlateauCdCr2O4_PRL,PRL2004_Penc,NPhys_Matsuda2007} %
In the case of HgCr$_2$O$_4$, previous x-ray and neutron diffraction measurements have detected changes in nearest neighbor bond length reflecting the bond order in the 1/2-magnetization plateau phase.\cite{NPhys_Matsuda2007} %
Although these extensive studies have established the importance of the spin-lattice coupling in the frustrated magnets, only a few experimental studies have directly observed the changes in exchange interactions in the bond-order states thus far.\cite{SKimura_HgCr2O4} %

In this paper, we report inelastic neutron scattering (INS) measurements on a triangular lattice antiferromagnet, which is a typical example of geometrically frustrated spin systems, CuFeO$_2$ (CFO) under applied magnetic fields of up to 13.4 T. %
We have observed spin-wave spectra in a magnetic field induced 1/5-magnetization plateau phase, which appears above $\sim$ 12.5 T below $\sim$ 10 K. %
We have determined the exchange interactions in the 1/5-magnetization plateau phase by calculating the spin-wave spectra using a Hamiltonian including the spin-lattice coupling effects for nearest neighbor (NN) exchange interactions. %
As a result, we have revealed that CFO exhibits a spin-driven bond order in the 1/5-magnetization plateau phase. %

\begin{figure*}[t]
\begin{center}
	\includegraphics[clip,keepaspectratio,width=16cm]{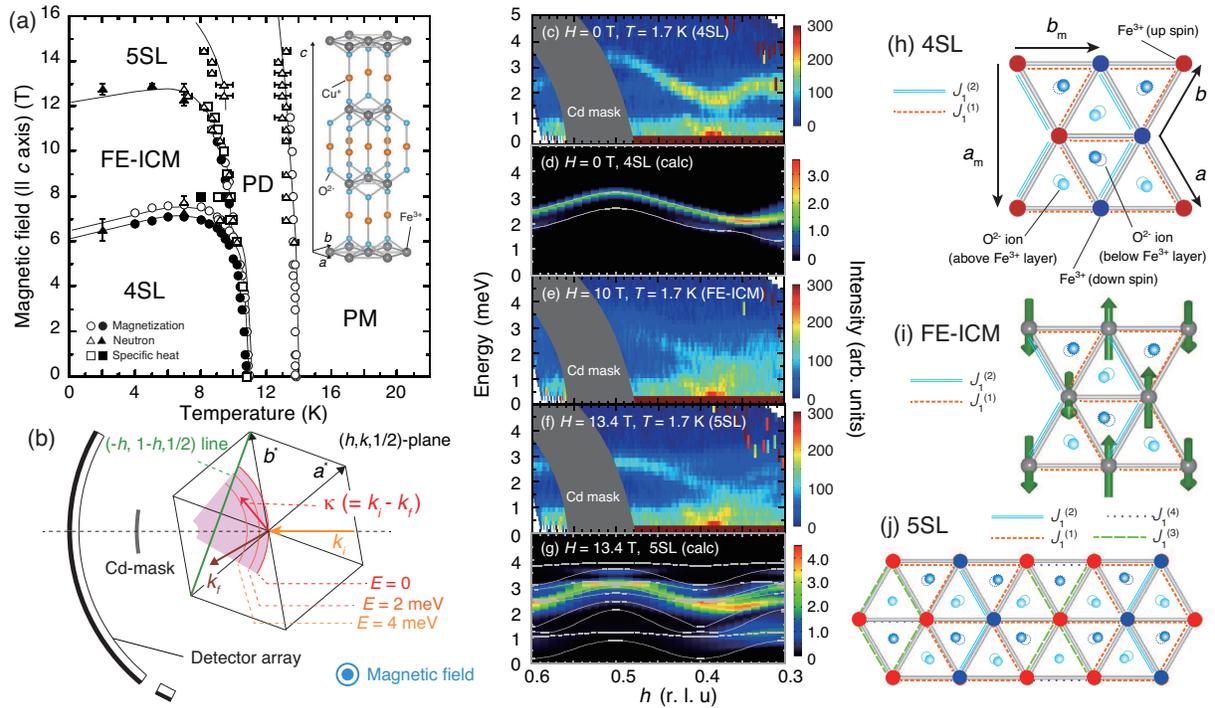}
	\caption{(Color Online) (a) The $H$-$T$ magnetic phase diagram of CFO (redrawn from Ref. \onlinecite{Mitsuda_2000}). %
	Open and filled symbols denote the transition temperatures (or fields) observed with increasing and decreasing temperature (or field), respectively. %
	(b) A schematic of the experimental setup for the present measurements (top view). %
	The $Q$-$E$ space measured with $E_i=8.6$ meV is shown by a pink shaded area. %
	 The observed INS spectra in (c) zero field, (e) under applied field of 10 T, and (f) 13.4 T. %
	 The calculated INS spectra for (d) the 4SL phase $(H=0$ T) and (g) the 5SL phase ($H=13.4$ T). %
	 The gray solid lines show the calculated spin-wave dispersion relation on the $(-h,1-h,1/2)$ line. %
	 [(h)-(j)] Schematics showing the oxygen displacements and the splitting of the NN exchange interactions with the magnetic structures in the (h) 4SL, (i) FE-ICM, and (j) 5SL phases. 
	$a_{\rm m}$ and $b_{\rm m}$ denote the monoclinic basis. }
	\label{8.6meV}
\end{center}
\end{figure*}

CFO has been extensively investigated as a geometrically frustrated magnet from 1990s.\cite{Mitsuda_1991,Mekata_1993} %
The crystal structure of CFO is shown in the inset of Fig. \ref{8.6meV}(a); the magnetic Fe$^{3+}$ ions are arranged in equilateral triangular lattice layers, which are separated by O$^{2-}$-Cu$^+$-O$^{2-}$ dumbbells. %
The Curie-Weiss temperature of this system has been estimated to be $\Theta_{CW}\sim -100$ K.\cite{Petrenko_2000,Mekata_1993}  %
On the other hand, CFO undergoes a magnetic phase transition from the paramagnetic (PM) phase to an incommensurate collinear magnetic phase, which is referred to as the partially disordered (PD) phase\cite{Mitsuda_1994_PD},  at $T_{\rm N1} = 14$ K in zero magnetic field. %
The large difference between $T_{\rm N1}$ and $|\Theta_{CW}|$ indicates the existence of the strong spin frustration in this system. 
The magnetic phase transition at $T_{\rm N1}$ is accompanied by a structural transition from the original trigonal structure to a monoclinic structure,\cite{Terada_CuFeO2_Xray,Ye_CuFeO2} so that the geometrical frustration is partly relieved. %
With further decreasing temperature from $T_{\rm N1}$, the system enters a collinear four-sublattice (4SL) antiferromagnetic ground state at $T_{\rm N2}\sim 11.2$ K. %
The  spin arrangement in the 4SL phase is shown in Fig. \ref{8.6meV}(h). %
When a magnetic field is applied along the $c$ axis at low temperatures, CFO exhibits successive magnetic phase transitions, as shown in the $H$-$T$ magnetic phase diagram in Fig. \ref{8.6meV}(a). %
The first-field-induced phase is referred to as the ferroelectric incommensurate-magnetic (FE-ICM) phase. %
The magnetic structure in this phase is a distorted screw-type structure,\cite{CompHelicity,Haraldsen_SW_FEphase} which breaks the inversion symmetry of the system and accounts for the ferroelectricity in this phase.\cite{Kimura_CuFeO2} %
The second-field-induced phase is the 1/5-magnetization plateau phase. %
Figure \ref{8.6meV}(j) shows the spin arrangement on a triangular lattice layer in this phase.\cite{Mitsuda_2000} %
Because the magnetic unit cell has five spins on its basal plane, this phase has been referred to as the five sublattice (5SL) phase. 

Interestingly, these magnetic-field-induced phase transitions are accompanied by distinct changes in crystal structure.\cite{Terada_CFO_Pulse,Terada_14.5T} %
This indicates that the spin-lattice coupling effect plays an important role for the field-induced transitions in CFO.\cite{Vishwanath_PRL_spin-phonon} %
Terada \textit{et al.} have explained the field-induced lattice deformations in terms of changes in Fe-O-Fe bonding angle.\cite{Terada_CFO_Pulse,Terada_LatticeModulation} %
They have pointed out that the antiferromagnetic (AF) NN interaction is enhanced by increasing the bonding angle, and vice versa.\cite{Terada_LatticeModulation} %
As a result, positions of the Fe$^{3+}$ and O$^{2-}$ ions are shifted so as to lower the  exchange energy in each of the magnetically ordered phases. %
Moreover, in CFO, an O$^{2-}$ ion is surrounded by three Fe$^{3+}$ ions, and thus a displacement of an O$^{2-}$ ion can affect three Fe-O-Fe bonds. %
This situation can lead to a variety of bond-order states associated with the magnetic orderings. %

Recently, we have identified the bond order in the 4SL phase.\cite{SingleDomainSW2} %
By means of INS measurements using a single crystal of CFO, we have revealed that the NN exchange interaction, $J_1$, splits into two interactions of $J_1^{(1)}$ and $J_1^{(2)}$; $J_1^{(1)}$ is a strong AF interaction connecting antiferromagnetically coupled NN spins and $J_1^{(2)}$ is a weak AF interaction connecting ferromagnetically coupled NN spins, as shown in Fig. \ref{8.6meV}(h). %
The atomic displacements associated with this bond order result in doubling of the unit cell along the [110] direction, %
which is consistent with the previous x-ray diffraction results.\cite{Terada_14.5T,Terada_CuFeO2_Xray,Ye_CuFeO2} %
A similar bond order and resulting atomic displacements were also found in the FE-ICM phase (see Fig. \ref{8.6meV}(i) and Refs.\onlinecite{SingleDomain_SW_CFGO,Haraldsen_SW_FEphase,Terada_14.5T,CFAO_xray}). %
On the other hand, in the 5SL phase, the changes in exchange interactions were not directly observed because of the difficulty of measuring the spin-wave spectra under high magnetic fields. %
In order to understand the role of the spin-lattice coupling effect in the field-induced transitions in this system, however, it is indispensable to determine the exchange interactions in the 5SL phase. %
In the present study, we have thus performed INS measurements on CFO under applied magnetic fields of up to 13.4 T. %

\section{EXPERIMENT}

The neutron scattering experiment was carried out using the chopper spectrometer LET at the ISIS spallation neutron source.\cite{LET} %
The detector coverage used in this experiment was from $-30^{\circ}$ to $50^{\circ}$. %
We used a vertical field superconducting cryomagnet whose maximum field is 13.5 T. %
The vertical open-angle of the magnet is from $-10^{\circ}$ to $15^{\circ}$. %
A number of incident energies ($E_i$) were selected by the multi-$E_i$ method. %
In the present analysis, we mainly used data measured with $E_i =$ 3.6 and 8.6 meV, for which the energy resolutions are estimated to be 0.049 and 0.17 meV, respectively. %
A single crystal of CFO was grown by the floating zone method,\cite{Zhao_FZ} and was cut into a plate shape with dimensions of 22, 5.0 and 3.7 mm for [001], [110] and $[1\bar{1}0]$ direction, respectively. %
During the experiment, we applied uniaxial pressure of $\sim 5$ MPa on the $[1\bar{1}0]$ surfaces of the sample using a duralumin clamp. %
This is because CFO has three magnetic domains in the magnetically ordered phases owing to the threefold rotational symmetry about the $c$ axis, and volume fractions of the three domains can be controlled by a small uniaxial pressure applied in the triangular lattice plane.\cite{SingleDomainSW} %
The sample with the uniaxial-pressure clamp was mounted in the cryomagnet so that the $c$ axis is parallel to the magnetic field. %
By measuring elastic magnetic Bragg reflections in the 5SL phase, we have found that the magnetic domain having the magnetic modulation wave vector parallel to the [110] direction has the volume fraction of 64 \%, and dominates over the other two domains (14\% and 22\%). %
The huge combined data set were handled by the HORACE software of ISIS.\cite{HORACE}

\begin{figure}[t]
\begin{center}
	\includegraphics[clip,keepaspectratio,width=8.5cm]{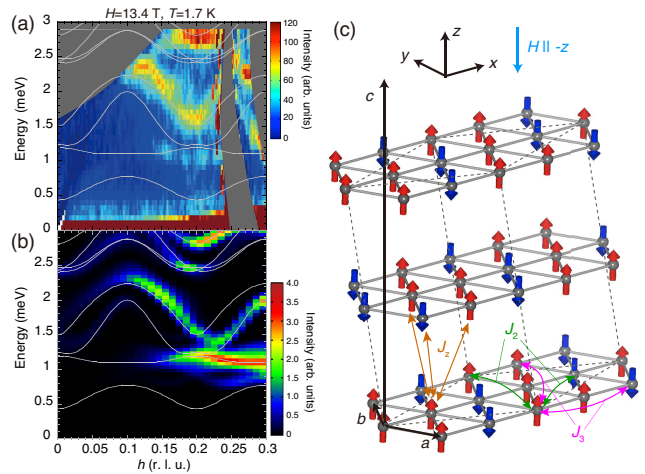}
	\caption{(Color Online) The (a) observed and (b) calculated spin wave excitation spectra along the $(h,h,0)$ line in the 5SL phase. %
	The calculated spin-wave dispersion relations are shown by solid gray lines. %
	(c) The magnetic structure in the 5SL phase and the definitions of $J_2$, $J_3$ and $J_z$. %
	Because of the minus sign between the spin angular momentum and the magnetic moment, the direction of the magnetic field is taken to be $-z$ direction. %
	The gray dashed lines show the magnetic unit cell for the 5SL phase. %
	 }
	\label{3.6meV}
\end{center}
\end{figure}

\section{RESULTS AND DISCUSSIONS}

Figures \ref{8.6meV}(c), \ref{8.6meV}(e), and \ref{8.6meV}(f) show the INS spectra measured with $E_i = 8.6$ meV, in the 4SL ($H=0$ T), FE-ICM (10 T), and 5SL (13.4 T) phases, respectively. %
These spectra were measured by setting the angle between the (110) direction of the crystal and the direction of the incident neutrons to $68^{\circ}\sim 73^{\circ}$, as shown in Fig. \ref{8.6meV}(b). %
The observed data were projected on to the $(-h,1-h,1/2)$ line, on which magnetic Bragg reflections appear in all the three phases.  %
In zero magnetic field, we found a distinct spin-wave branch lying in the energy range of $E = 2\sim 3.5$ meV, as shown in Fig. \ref{8.6meV}(c). %
This corresponds to the higher energy branch observed in the previous INS measurements in the 4SL phase.\cite{SingleDomainSW,SingleDomainSW2} %
This spin-wave spectrum also indicates that the inelastic scattering signals from the two minority domains were negligible in the present experiment, %
because the observed spectrum is almost the same as that in the nearly `single-domain' 4SL phase in Ref. \onlinecite{SingleDomainSW}. %
In Fig. \ref{8.6meV}(d), we show a calculated INS spectrum along the $(-h,1-h,\frac{1}{2})$ line in the single-domain 4SL phase, which also qualitatively agrees with the present result. %
In the FE-ICM phase, the excitation spectrum became rather diffusive, as shown in Fig. \ref{8.6meV}(e). %
This is characteristic of the incommensurate and noncollinear magnetic ordering in the FE-ICM phase, and is similar to the magnetic excitation spectra observed in the FE-ICM phase of CuFe$_{1-x}$Ga$_x$O$_2$ $(x=0.035)$.\cite{SingleDomain_SW_CFGO,Haraldsen_SW_FEphase} %
In the 5SL phase, we found that distinct spin-wave branches were retrieved, as shown in Fig. \ref{8.6meV}(f). %

\begin{table*}[t]
\begin{center}
\begin{tabular}{ccccccccccc}
\hline
\hline
magnetic phase &sample         & $J_1^{(1)}$ & $J_1^{(2)}$ & $J_1^{(3)}$ & $J_1^{(4)}$ & $J_2$      & $J_3$     & $J_{z}$    & $D$ & Ref. \\
\hline
4SL  ($H=0$ T)                 &CuFeO$_2$     & $ -0.176$    & $-0.060$     & -                   & -                  & $-0.041$ & $-0.142$ & $-0.071$ & $0.064$ & \onlinecite{SingleDomainSW2}\\
FE-ICM  ($H=0$ T)   &CuFe$_{1-x}$Ga$_x$O$_2$ ($x=0.035)$ & $ -0.169$    & $-0.066$     & -                   & -                  & $-0.070$ & $-0.098$ & $-0.070$ & $0.014$ & \onlinecite{SingleDomain_SW_CFGO}\\
5SL  ($H=13.4$ T)                    &CuFeO$_2$ &  $ -0.18$    & $-0.06$     & $-0.10$      & $-0.14$     & $-0.06$ & $-0.15$ & $-0.06$  & $0.064$ & This work\\
\hline
\hline
\end{tabular}
\caption{The Hamiltonian parameters in the 4SL phase (from Ref. \onlinecite{SingleDomainSW2}),\cite{comment1} the FE-ICM phase (from Ref. \onlinecite{SingleDomain_SW_CFGO}) and the 5SL phase (in meV). %
 }
\label{Table_J}
\end{center}
\end{table*}

In Fig. \ref{3.6meV}(a), we show the spin-wave excitation spectrum along the $(h,h,0)$ line measured with $E_i=3.6$ meV at $H=13.4$ T and $T=1.7$ K. %
The observed spectrum seems to be consistent with a theoretical prediction by Haraldsen \textit{et al}, in which the spin-wave excitations under applied magnetic fields were calculated using Monte Carlo simulations and variational method for two dimensional triangular lattice.\cite{Haraldsen_CFO_Prediction} %
However, in CFO, exchange interactions between adjacent triangular lattice layers are quite important as was pointed out in the previous theoretical study by Fishman \textit{et al}.\cite{Fishman_Stacking_PRB} %
In the present study, we have thus employed the three dimensional magnetic structure of the 5SL phase to calculate the spin-wave spectra. %

Figure \ref{3.6meV}(c) shows the magnetic structure in the 5SL phase.\cite{Mitsuda_2000} %
The magnetic unit cell in the 5SL phase contains two triangular lattice layers, each of which has five spins, and therefore, the 5SL phase actually has ten sublattices.\cite{Fishman_Stacking_PRB} %
To calculate the spin-wave spectrum in the 5SL phase, we have employed a conventional linear spin-wave theory with a Hamiltonian, %
\begin{eqnarray}
\mathcal{H}=-\frac{1}{2}\sum_{i\ne j}J_{ij}\mbox{\boldmath $S$}_i\cdot \mbox{\boldmath $S$}_j-D\sum_i(\mbox{\boldmath $S$}_i^z)^2+ g\mu_B\sum_i\mbox{\boldmath $S$}_i^zH,
\label{Hamiltonian}
\end{eqnarray}
where  $S=5/2$, and $D$ is a uniaxial anisotropy. The gyromagnetic ratio, $g$, is assumed to be 2. $H$ and $\mu_B$ are the applied magnetic field and the Bohr magneton, respectively. %
Similarly to our previous work on the spin-wave excitations in the 4SL phase,\cite{SingleDomainSW2} nearest, second, and third neighbor exchange interactions within the triangular lattice layers ($J_1, J_2$ and $J_3,$) and an exchange interaction between the adjacent layers ($J_z$) are employed in Eq. (\ref{Hamiltonian}). %
As for $J_1$, we have introduced the spin-lattice coupling effect in the same spirit as the previous works on the 4SL and FE-ICM phases.\cite{SingleDomainSW2,SingleDomain_SW_CFGO,Haraldsen_SW_FEphase} %
The displacements of each O$^{2-}$ ion are assumed from the spin arrangements of the three neighboring Fe$^{3+}$ ions. %
Hereby, $J_1$ splits into four different NN interactions, $J_1^{(1)}, J_1^{(2)}, J_1^{(3)}$ and $J_1^{(4)}$ as shown in Fig. \ref{8.6meV}(j). %

The procedure of the spin-wave calculation for the 5SL phase is essentially the same as those in the previous works on the 4SL phase.\cite{SingleDomainSW2,Fishman_SW} %
We have applied a Holstein-Primakoff $1/S$ expansion about the classical limit to the Hamiltonian of Eq.(\ref{Hamiltonian}). %
We express the spins, $\mbox{\boldmath $S$}_i$, on each sublattice using Fourier transformed boson operators. %
By solving the Heisenberg equation of motion for the boson operators, the spin-wave dispersion relations and the INS cross section were calculated. %
To obtain the resolution-convoluted neutron scattering spectra, the experimental resolutions for $E_i=3.6$ and 8.6 meV were taken into account. %
The Hamiltonian parameters were adjusted so that the calculations reproduce the observed data, and finally, the best fit was obtained for the parameters shown in Table \ref{Table_J}.\cite{comment3} %
The calculated spectra for the $(h,h,0)$ and $(-h,1-h,\frac{1}{2})$ lines with $E_i=3.6$ and 8.6 meV are shown in Fig. \ref{3.6meV}(b) and \ref{8.6meV}(g), respectively. %

Comparing the parameters for the 5SL phase with those in the 4SL and FE-ICM phases, we have found that $J_1^{(1)}$ and $J_1^{(2)}$ are nearly common in all the three phases. %
This is consistent with the fact that the spin arrangements and the oxygen displacements associated with $J_1^{(1)}$ and $J_1^{(2)}$ in the 5SL phase are the same as those in the 4SL phase, as shown in Figs. \ref{8.6meV}(h) and \ref{8.6meV}(j). %
As for $J_1^{(3)}$ and $J_1^{(4)}$, their magnitudes are found to be smaller than that of $J_1^{(1)}$. %
This is also reasonable because they connect two ferromagnetically coupled spins. %
In addition, the exchange paths of $J_1^{(3)}$ and $J_1^{(4)}$ include the O$^{2-}$ ions surrounded by three up spins, and the Fe-O-Fe bonding angles between the three up spins are expected to remain the same as each other. %
Therefore, the `reductions' in magnitudes of $J_1^{(3)}$ and $J_1^{(4)}$ are expected to be smaller than that of $J_1^{(2)}$. %
We have found that the experimentally determined parameters are in good agreement with this scenario. %
This is the direct evidence for the spin-driven bond order in the 5SL phase. %
Moreover, the present results have also demonstrated that the field-induced phase transition from the FE-ICM phase to the 5SL phase is accompanied by the bond-order transition. %

In contrast to the drastic changes in $J_1$, the spin-lattice coupling effects on the distant interactions ($J_2, J_3, J_z$) are relatively small,\cite{SingleDomainSW2} and these interactions in the 5SL phase are comparable to those in the 4SL and FE-ICM phases. %
This might be because a direct exchange interaction between the NN Fe$^{3+}$ ions, which is assumed to be ferromagnetic,\cite{Mekata_1993} competes with the Fe-O-Fe superexchange interaction, and therefore $J_1$ is highly sensitive to the small lattice distortions as compared to $J_2$, $J_3$ and $J_z$. %

In summary, we have investigated the spin-wave excitations and the spin-lattice coupling in the 5SL phase of CFO, by means of the INS measurements under applied field of 13.4 T. %
Comparing the observed spin-wave spectra with the calculations including the spin-lattice coupling effects for the NN exchange interactions, we have revealed that CFO exhibits the spin-driven bond order in the 5SL phase. %
It should be emphasized that we have constructed the model of the bond order by taking into account the fact that an O$^{2-}$ ion belongs to three Fe-O-Fe bonds in CFO. %
The present results suggest the importance of topology of exchange-interactions paths for understanding the exotic spin-lattice coupling phenomena, specifically spin-driven bond order, in geometrically frustrated magnets. %

The inelastic neutron scattering experiment at ISIS was carried out along the proposal No. BR1220149. %
We are grateful to Dr. T. Guidi for technical support in the experiment. %
This work was supported by Grants-in-Aid for Young Scientist (B) (Grant Nos. 23740277 and 25800203). %
N.T. is supported by the JSPS Postdoctoral Fellowships for Research Abroad. %
The images of the crystal and magnetic structures in this paper were depicted using the software VESTA\cite{VESTA} developed by K. Monma.

\bibliography{main}

\begin{thebibliography}{34}
\bibitem[{com({\natexlab{a}})}]{comment3}
\bibinfo{note}{In the present analysis, the Hamiltonian parameters were
  adjusted manually, and therefore, we could not determine the errors of each
  parameter. Through trial and error, we roughly estimate that the errors are
  in the order of $\sim 0.01$ meV.}
\bibitem[{com({\natexlab{b}})}]{comment1}
\bibinfo{note}{In Ref. \onlinecite{SingleDomainSW2}, we have assumed that
  $J_1^{(1)}, J_2, J_3$ and $J_z$ also split into two different interactions,
  but the splittings are fairly small as compared with the difference between
  $J_1^{(1)}$ and $J_1^{(2)}$. Therefore, in Table \ref{Table_J}, the small
  splittings are neglected and we have shown the averaged values.}
\end{thebibliography}

\end{document}